\documentclass[11pt]{article}
\usepackage{hyperref}
\pdfoutput=1

\begin{document}
\title{Bouncing Water Droplet on a Superhydrophobic Carbon Nanotube Array}
\author{Adrianus I. Aria and Morteza Gharib \\
\\\vspace{6pt} Graduate Aeronautical Laboratories, \\ California Institute of Technology, Pasadena, CA 91125, USA}
\maketitle

\begin{abstract}
Over the past few decades, superhydrophobic materials have attaracted a lot of interests, due to their numerous practical applications. Among various superhydrophobic materials, carbon nanotube arrays have gained enormous attentions simply because of their outstanding properties.  The impact dynamics of water droplet on a superhydrophobic carbon nanotube array is shown in this fluid dynamics video. 

\end{abstract}


\section{Video description}

The first two parts of the video show the impact dynamic of 30 $\mu$l water droplet at different impact velocity. At low impact velocity of 1.03 m/s, the water droplet deforms upon impact and eventually bounces off completely of the surface of the array. At higher impact velocity of 2.21 m/s, the droplet breaks up into many smaller droplets and eventually bounces off completely of the surface of the array.

The coefficient of restitution of water droplet at very low impact velocity can be seen clearly by dropping a water droplet on a slightly tilted carbon nanotube array. At tilt angle of 2.5 $^\circ  $ the droplet skips off of the surface of the array multiple times without showing any sign of pinning on the surface of the array, as demonstrated in the third part of the video. The fourth part of the video shows the sliding/rolling behavior of the droplet along the surface of a U-shaped carbon nanotube array. The fifth part of the video shows the impact of two identical 14  $\mu$l water droplets to one another on a U-shaped carbon nanotube arrays. Upon impact, these two water droplets, which come from the opposite direction, merge to form one larger droplet. 

All parts of the video were captured by high speed camera operated at various frame rates. The droplet was illuminated from behind with a diffuse halogen light. The droplet was dropped on the surface by flat-tipped needle and the volume of the droplet was controlled precisely by syringe pump.
\\\\
This work was supported by The Charyk Foundation and The Fletcher Jones Foundation. 

\end{document}